\documentclass[
 reprint,
superscriptaddress,
 amsmath,amssymb,
 aps, prb
]{revtex4-2}
\usepackage[utf8]{inputenc}
\usepackage{multirow}
\usepackage{braket}
\usepackage[colorlinks=true,allcolors=blue]{hyperref}
\usepackage{mathtools}
\usepackage{babel}
\usepackage{graphicx}
\usepackage{dcolumn}
\usepackage{bm}
\usepackage[capitalise]{cleveref}
\usepackage{xcolor}
\usepackage{xspace}

\newcommand*{\gowo}{\ensuremath{G_0W_0}\xspace}
\newcommand{\gw}{\ensuremath{GW}\xspace}
\newcommand{\qsgw}{qp$GW$-I\xspace}
\newcommand{\qpgw}{qp$GW$-II\xspace}
\newcommand{\scgw}{sc$GW$\xspace}

\newcolumntype{P}[1]{>{\centering\arraybackslash}p{#1}}

\newlength{\templen}

\begin{document}
\author{Gaurav Harsha}
\author{Vibin Abraham}
\author{Ming Wen}
\affiliation{Department of Chemistry, University of Michigan, Ann Arbor, Michigan 48109, USA}
\author{Dominika Zgid}
\affiliation{Department of Chemistry, University of Michigan, Ann Arbor, Michigan 48109, USA}
\affiliation{Department of Physics, University of Michigan, Ann Arbor, Michigan 48109, USA}
\title{Quasiparticle and fully self-consistent \gw methods: an unbiased analysis\\using Gaussian orbitals}

\begin{abstract}
We present a comparison of various approximations to self-consistency in the $GW$ method, including the one-shot $G_0 W_0$ method, different quasiparticle self-consistency schemes, and the fully self-consistent $GW$ (\scgw) approach. To ensure an unbiased and equitable comparison, we have implemented all the schemes with the same underlying Matsubara formalism, while employing Gaussian orbitals to describe the system. 
Aiming to assess and compare different $GW$ schemes, we analyze band gaps in semiconductors and insulators, as well as ionization potentials in molecules.
Our findings reveal that for solids, the different self-consistency schemes perform very similarly. However, for molecules, full self-consistency outperforms all other approximations, i.e., the one-shot and quasiparticle self-consistency \gw schemes.
Our work highlights the importance of implementation details when comparing different $GW$ methods.
By employing state-of-the-art fully self-consistent, finite temperature $GW$ calculations, we have successfully addressed discrepancies in the existing literature regarding its performance.
Our results also indicate that when stringent thresholds are imposed, the \scgw method consistently yields accurate results.
\end{abstract}
\maketitle

\section{Introduction}
The $GW$ approximation~\cite{hedin_new_1965,aryasetiawan_thegwmethod_1998,reining_gw_2018,golze_gw_2019} is ubiquitously used in chemistry and physics to study correlated quantum many-body systems beyond the mean-field methods such as the density functional theory~\cite{kohn_ground-state_1960, kohn_self-consistent_1965} (DFT) and Hartree-Fock (HF).
$GW$ is one of the simplest in a series of systematically improvable approximations to Hedin's formulation of the many-body perturbation theory~\cite{hedin_new_1965}, in which the self-energy is approximated as the product of single-particle Green's function $G$ and the screened Coulomb interaction $W$.
It has proven to be fairly accurate in describing photo-electron spectra in both finite~\cite{stan_fully_2006,caruso_unified_2012,marom_benchmark_2012,van_setten_gw-method_2013,caruso_benchmark_2016,maggio_gw_2017,forster_low-order_2020,forster_low-order_2021,wen_comparing_2024,abraham_relativistic_2024} and extended systems~\cite{faleev_all-electron_2004,shishkin_self-consistent_2007,kutepov_ground-state_2009,bruneval_quasiparticle_2014,kutepov_ground_2017,grumet_beyond_2018,pokhilko_broken-symmetry_2022,pokhilko_evaluation_2023}, and is also considered as an ideal starting point for sophisticated quantum embedding theories such as dynamical mean-field theory~\cite{bierman_gw+dmft,kotliar_electronic_2006,choi_first_2016,lee_diatomic_2017,nilsson_multitier_2017,zhu_abinitio_2021} and self-energy embedding theory~\cite{lan_testing_2017,Iskakov_MnO_NiO,Yeh_SrMnO,SEET_periodic}.
The success of $GW$ can be attributed to its ability to accurately describe the screening of Coulomb interaction between electrons in contrast to fully iterative methods such as self-consistent second order (GF2)~\cite{PhysRevB.100.085112} based on the diagrammatic expansion with bare Coulomb interactions. 
Moreover, $GW$ also accounts for electron exchange, albeit at the lowest order, i.e., using the Fock diagram.
Diagrammatic similarities between \gw and other accurate methods such as coupled cluster (CC) and random phase approximation (RPA) are also well known, which further explain its accuracy~\cite{scuseria_particle-particle_2013,lange_relation_2018,quintero-monsebaiz_connections_2022,tolle_exact_2023}.

The formal computational scaling of \gw equations is $\mathcal{O}(N^6)$, where $N$ is a measure of the system size.
After approximations such as the decomposition of two-electron integrals~\cite{vahtras_integral_1993,sun_gaussian_2017,ye_fast_2021}, this scaling is brought down to $\mathcal{O}(N^4)$ making \gw an attractive method to study large, realistic systems.
By using tensor hyper-contraction techniques, this computational scaling can be reduced even further to $\mathcal{O}(N^3)$~\cite{yeh_low-scaling_2024}.

Despite the low computational scaling, storing and manipulating dynamical quantities such as the Green's function and the self-energy can still become costly.
As a result, early implementations~\cite{hybertsen_ab_1987,godby_metal-insulator_1989,oschlies_gw_1995} of the $GW$ method relied on numerous approximations, e.g., plasmon-pole models for the dielectric function~\cite{engel_generalized_1993,larson_role_2013}, in addition to the usual ones, such as the use of pseudopotentials and finite basis-sets for electronic orbitals.
The notion of one-shot $GW$ (or $\gowo$), where just a single iteration of $GW$ is performed, was also introduced.
The diagonal approximation is another one that is now commonly used with $\gowo$, in which corrections are introduced only for the quasiparticle energies while eigenvectors of the initial mean-field calculation are retained. 
This results in a self-energy with no off-diagonal terms (in the basis of mean-field eigenvectors).

It is clear that performing only a single iteration of $GW$ leads to results that depend strongly on the mean-field starting point.
However, with sufficient experience, an optimal reference for \gowo can sometimes be found~\cite{korzdorfer_strategy_2012,bruneval_benchmarking_2013}.
For example, in semiconductors and insulators, $\gowo$ based on DFT calculations with local density approximation (LDA) or Predew-Becke-Ernzerhof~\cite{perdew_generalized_1996} (PBE) functional provides accurate band gaps~\cite{aryasetiawan_thegwmethod_1998,onida_electronic_2002,rinke_combining_2005,fuchs_quasiparticle_2007,chen_accurate_2015,gant_optimally_2022,rodrigues_pela_critical_2024}, while for molecules, $\gowo$ with hybrid DFT reference, containing some level of exact exchange, provides accurate ionization potentials (IP) ~\cite{marom_benchmark_2012,zhang_recommendation_2022,li_benchmark_2022,forster_two-component_2023,wen_comparing_2024,abraham_relativistic_2024}.
In fact, such a $\gowo$ description can be tuned to perform better than full self-consistency.

As algorithmic and computational capabilities improved over time, implementations of fully self-consistent solution of $GW$ equations, called as \scgw, have become more common~\cite{holm_fully_1998, faleev_all-electron_2004, stan_fully_2006, shishkin_self-consistent_2007, caruso_unified_2012, koval_fully_2014, cao_fully_2017, grumet_beyond_2018, ren_all-electron_2021, yeh_relativistic_2022, yeh_fully_2022}.
Self-consistency guarantees black-box results, independent of the mean-field reference.
Moreover, at self-consistency, the $GW$ approximation is derivable from the Luttinger-Ward functional~\cite{luttinger_ground-state_1960,baym_conservation_1961,hyrkas_cutting_2022} and, therefore, ensures that thermodynamic variables such as total energy, entropy, grand-potential, etc., are well defined, i.e., different ways of calculating these quantities yield identical results.

\begin{figure}
    \centering
    \includegraphics[width=1.0\linewidth]{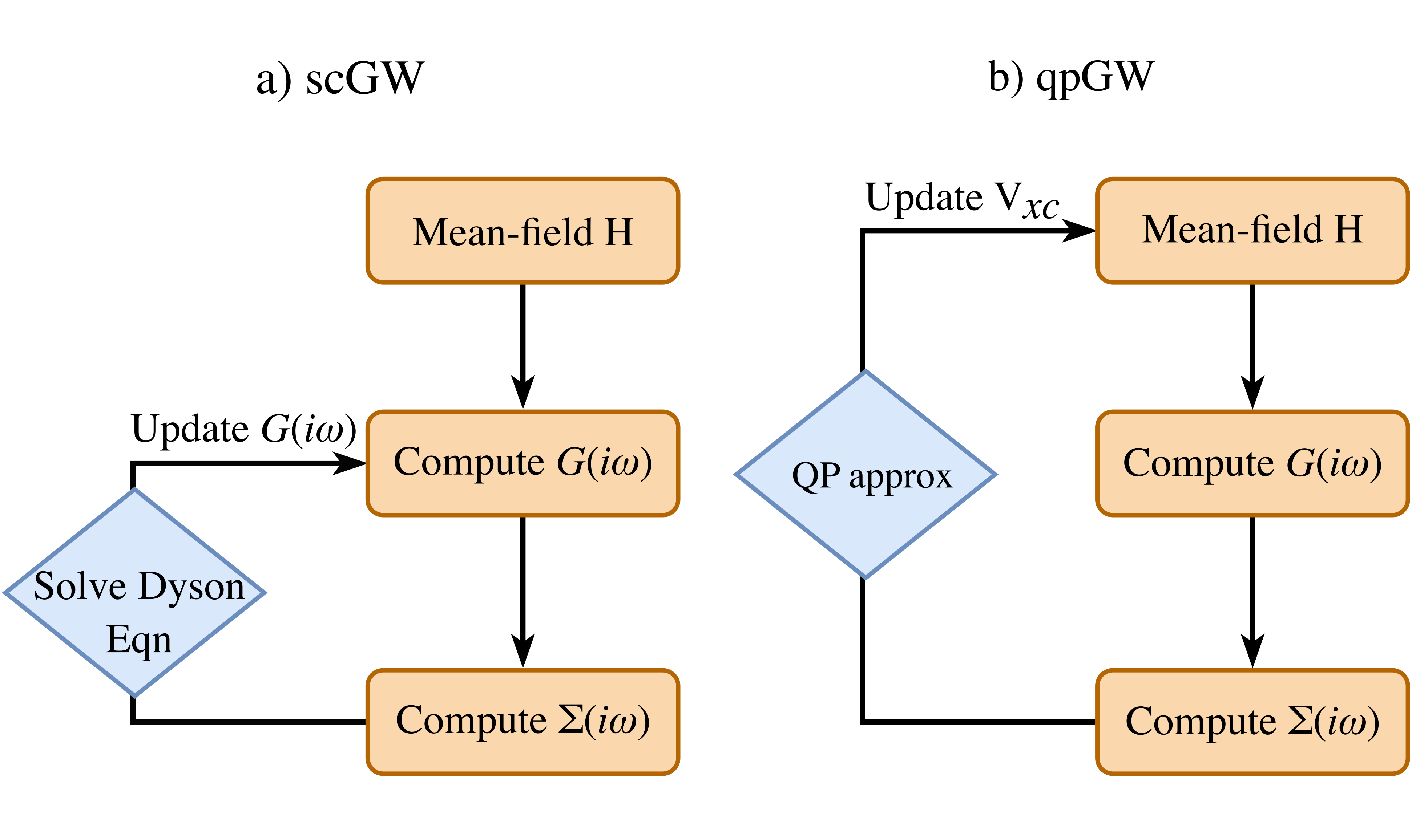}
    \caption{Schematic representation of different self-consistency schemes in the \gw approximation. The central object in \scgw is the correlated Green's function $G$, whereas in qp\gw, it is the effective mean-field Hamiltonian.} 
    \label{fig:gw_schemes}
\end{figure}

In the pursuit of combining the merits of \scgw and $\gowo$, two notable methodologies for quasiparticle self-consistent $GW$ have emerged~\cite{vanschilfgaarde_quasiparticle_2006,kotani_quasiparticle_2007,kutepov_electronic_2012,kutepov_linearized_2017}.
The first approach by van Schilfgaarde et al.~\cite{vanschilfgaarde_quasiparticle_2006,kotani_quasiparticle_2007}, hereafter referred as \qsgw, involves the construction of an effective, static (or frequency-independent) exchange-correlation potential from the dynamical (or frequency-dependent) self-energy in each iteration of the $GW$ cycle.
This effective potential is then used to solve the quasiparticle equation yielding a new Green's function for the next iteration.
Another approach by Kutepov et al.~\cite{kutepov_electronic_2012}, referred here as \qpgw, proposes a first-order linear-in-frequency approximation to the self-energy.
Through further manipulations, an effective Hamiltonian is derived, with its eigenvalues corresponding to quasiparticle energies. 
Schematic diagrams summarizing the quasiparticle self-consistency schemes and \scgw are shown in \cref{fig:gw_schemes}.

Both forms of quasiparticle self-consistency, \qsgw and \qpgw, have been shown to be reasonably accurate both for band gaps in semiconductors~\cite{faleev_all-electron_2004, kutepov_electronic_2012, kaplan_quasi-particle_2016}, as well for ionization potential in molecules~\cite{caruso_benchmark_2016}.
By their construction, they also remedy the problem of mean-field-dependence present in $\gowo$.
Several other ways to marry the benefits of $\gowo$ and \scgw have also been proposed, e.g., partial self-consistency in $GW_0$~\cite{holm_fully_1998,garcia-gonzalez_self-consistent_2001}, eigenvalue self-consistent $GW$ (ev$GW$)~\cite{hybertsen_electron_1986}, etc., most of which only partially cure the reference-dependence.
For the purpose of this paper, we will focus on one-shot, quasiparticle, and fully self-consistent $GW$ methods.

A straightforward comparison of new approximations, such as quasiparticle self-consistency, in $GW$ with existing methodologies of $\gowo$ and particularly \scgw is difficult and presents several challenges. 
First, accurate implementations of \scgw have become possible only recently.
The issue is further complicated by the fact that each $GW$ implementation varies in several technical aspects. These include, but are not limited to, the choice of single-particle basis, whether finite-temperature or zero-temperature formalism is employed, and accuracy of grids for time- and frequency-representation of Green's function and self-energy.
Furthermore, when finite-temperature formalism is employed, analytical continuation (AC) of the results from Matsubara to the real frequency axis becomes necessary.
This compels another decision in selecting an appropriate AC method from among many alternatives~\cite{vidberg_solving_1977, yoon_analytic_2018, fei_nevanlinna_2021, fei_analytical_2021, huang_robust_2023}.
Rigorous comparisons of different quasiparticle and full self-consistency schemes in $GW$ have therefore been rare~\cite{koval_fully_2014,caruso_benchmark_2016,grumet_beyond_2018,kutepov_ground_2017,kutepov_full_2022}.

Grumet et al. conducted a comparative analysis of different self-consistency schemes in $GW$, as outlined in Ref.~\onlinecite{grumet_beyond_2018}. Their study revealed that fully self-consistent \gw, performed on the imaginary-time grid, yielded larger band gaps compared to the fully self-consistent \gw reported by Kutepov \cite{kutepov_self-consistent_2017}.
Additionally, the band gaps obtained using the \scgw approach by Kutepov were notably larger than those from the \qsgw scheme reported by Shilfgaarde-Kotani-Faleev~\cite{faleev_all-electron_2004,kotani_quasiparticle_2007}, which was also included for comparison purposes in Grumet's work.
Consequently, an unbiased determination of whether quasiparticle self-consistent \gw results closely resemble \scgw, as reported by Kutepov, or if there is a difference between \scgw and \qsgw, as reported by Grumet et al., was not feasible based on their findings. 
Even for molecules, the ambiguity in the relative accuracy of \qsgw and \scgw persists.
While Koval and co-authors~\cite{koval_fully_2014} reported that \scgw provides better IPs as compared to \qsgw, similar study by Caruso et al.~\cite{caruso_benchmark_2016} suggested otherwise.
Meanwhile, the application and analysis of \qpgw for molecules remain completely unexplored.

Our study aims to resolve this discrepancy by implementing and comparing $\gowo$, \qsgw, \qpgw, and \scgw methods, all within the framework of Matsubara Green's functions as explored in Ref.~\onlinecite{yeh_fully_2022}. By maintaining identical underlying implementation details across the various methods, we ensure a fair and unbiased comparison. 
Furthermore, our implementations utilize Gaussian orbitals as the one-particle basis, eliminating such bias as is present when methods implemented employing different one-particle bases, e.g., linearized augmented-plane-wave (LAPW) or plane wave basis, are compared against each other.
To differentiate between $\gowo$, \qsgw, \qpgw, and \scgw methods, we examine band gaps in selected semiconductors and insulators, as well as ionization potentials in selected molecules. We also compare trends and potential issues that arise in the convergence of self-consistent schemes.
While other studies comparing different approximations in the \gw method have been conducted in the literature, e.g., in Refs.~\cite{stan_levels_2009,caruso_benchmark_2016,lei_gaussian-based_2022}, they either utilize different implementations for \scgw and qp$GW$, or lack a direct comparison between the two methods.

\section{\label{sec:theory}Theory}
We first describe the theoretical details for the various $GW$ methodologies that we implement and compare in this paper.
In the finite-temperature Matsubara formalism employed here, the equilibrium single-particle Green's function on the imaginary-time axis is defined as
\begin{equation}
    G_{ij} (\tau) = -\frac{1}{\mathcal{Z}} \mathrm{Tr} \left[
        e^{-(\beta - \tau) (H - \mu N)} c_{i}
        e^{-\tau (H - \mu N)} c_{j}^{\dagger}
    \right],
    \label{eq:matsubara-gf}
\end{equation}
where $\tau \in [0, \beta)$ is the imaginary time, $H$ and $N$ are the Hamiltonian and number operators, respectively, $\mu$ is the chemical potential, and operators $c_i$ ($c_i^{\dagger}$) annihilate (create) a particle in the $i$th spin-orbital. Finally, the partition function $\mathcal{Z}$ is defined as
\begin{equation}
    \mathcal{Z} = \mathrm{Tr} \left[ e^{-\beta (H - \mu N)} \right],
\end{equation}
where $\beta = 1 / k_B T$ is the inverse temperature.
The Fourier transform between the imaginary-time and the Matsubara Green's function is defined as
\begin{subequations}
    \label{eq:fourier}
    \begin{align}
        G_{ij} (i\omega_n)
        &=
        \int_0^\beta d \tau G_{ij} (\tau) e^{i \omega_n \tau},
        \\
        G_{ij} (\tau)
        &=
        \frac{1}{\beta} \sum_n G_{ij} (i\omega_n) e^{-i\omega_n \tau}.
    \end{align}
\end{subequations}
The fermionic Matsubara frequencies take the values $\omega_n = (2n + 1) \pi / \beta$, where $n = 0, \pm 1, \pm 2$, and so forth.
Further, we use a general two-body electronic Hamiltonian, written as
\begin{equation}
    H = \sum_{ij} (H_0)_{ij} c_{i}^\dagger c_j
    + \frac{1}{2} \sum_{ijkl} U_{ijkl} c_i^\dagger c_k^\dagger c_l c_j,
\end{equation}
where the one-electron integrals $H_0$ describe the kinetic energy and the interaction of electrons with the nuclei and other external potential, while the two-electron integrals $U$ describe the Coulomb repulsion between electrons.
Generalizations of this Hamiltonian to the case of periodic / extended systems are straightforward. For an explicit notation and equations, the reader may refer to Ref.~\onlinecite{yeh_fully_2022}.
We should note that in all the discussion below, unless stated otherwise, we consider all the quantities, such as the Hamiltonian, Green's function, self-energy, etc. in the non-orthogonal atomic-orbital (AO) basis.

\subsection{\label{subsec:gw-approx}The $GW$ approximation}
Hedin's formulation of the many-body perturbation theory describes a coupled set of equations for the interacting single-particle Green's function $G$, the vertex function $\Gamma$, the irreducible polarization $\Pi$, the screened Coulomb interaction $W$, and the self-energy $\Sigma$ of a system.

In the simplest non-trivial approximation, higher order corrections to the vertex function are ignored, i.e., $\Gamma$ is defined as a Dirac-delta function.
As a result, the polarization $\Pi$ and the effective screened Coulomb interaction $W$ are given by
\begin{subequations}
    \label{eq:hedin-gw-polarization}
    \begin{align}
        \Pi_{abcd} (\tau) &= 
        G_{da} (\tau) G_{bc} (-\tau),
        \\
        {W}_{ijkl} (i \Omega_n) &=
        U_{ijkl} \nonumber
        \\
        & \, + \sum_{abcd} U_{ijab} \Pi_{abcd} (i \Omega_n) W_{cdkl} (i \Omega_n),
        \label{eq:screened-coul}
    \end{align}
\end{subequations}
where $\Omega_n = 2n\pi / \beta$ are the bosonic Matsubara frequencies with $n = 0, \pm 1, \pm 2, \cdots$.
Subsequently, the total \gw self-energy is expressed as a sum of static and dynamical terms, i.e.,
\begin{subequations}
    \label{eq:hedin-gw-selfenergy}
    \begin{align}
        \Sigma_{\mathrm{total}} (\tau) &=\Sigma_\infty + \Sigma (\tau),
        \\
        \Sigma_{\infty, ij} &=
        \sum_{ab} \rho_{ab} \left( U_{ijba} - U_{iabj} \right),
        \\
        \Sigma_{ij} (\tau) &=
        - \sum_{ab} G_{ab} (\tau) \tilde{W}_{iabj} (\tau),\label{eq:sig:gw1}
    \end{align}
\end{subequations}
where the static self-energy $\Sigma_{\infty}$ arises from the contraction of density matrix $\rho = - G(\tau = \beta^-)$ with the bare Coulomb interaction $U$ and contains the Hartree-Fock diagrams, while the dynamical contribution comes from the effective screened Coulomb interaction $\tilde{W}$ defined as
\begin{equation}
    \tilde{W} (\tau)
    \xleftrightarrow[\textrm{Transform}]{\textrm{Fourier}}
    \tilde{W} (i \Omega_n) = W (i \Omega_n) - U.
    \label{eq:screening-fourier}
\end{equation}
Finally, the full correlated Green's function is defined as
\begin{subequations}
    \label{eq:hedin-gw-dyson}
    \begin{align}
        [\mathbf{G}^{-1}] (i\omega_n) &=
         [\mathbf{G}^{-1}_0] (i\omega_n) -\mathbf{\Sigma} (i\omega_n), \label{eq:dyson}\\
        [\mathbf{G}^{-1}_0] (i\omega_n) &=  (i \omega_n + \mu) \mathbf{S} - \mathbf{F} 
    \end{align}
\end{subequations}
where $\mathbf{S}$ denotes the overlap matrix, $ \mathbf{\Sigma} (i\omega_n)$ is the Fourier transform of $\Sigma_{ij} (\tau)$, and $\mathbf{F=H_0 + \Sigma_{\infty}}$ is a Fock matrix.

\subsubsection{\label{subsubsec:scgw}Self-consistent $GW$}
In \scgw, \cref{eq:hedin-gw-polarization,eq:hedin-gw-selfenergy,eq:screening-fourier,eq:hedin-gw-dyson} are solved in an iterative manner until self-consistency is achieved.
Numerically, convergence means that updates from the subsequent iterations to the quantities involved, such as $G$ and $\Sigma$ or derived properties such as total energy, are smaller than a specified threshold.
We initialize each calculation with $\Sigma_{\infty}^{(0)} = V_{\mathrm{eff}}$ and $\Sigma^{(0)} (\tau) = 0$, where $V_{\mathrm{eff}}$ represents the effective Hartree and exchange potential obtained from the initial mean-field calculation.
This defines the initial Green's function through \cref{eq:dyson}.
Following that, the algorithm used for \scgw in this work can be succinctly described as:
\begin{enumerate}
    \item At the $n$th iteration, use the Green's function $\mathbf{G}^{(n)}$ to build the static and dynamical parts of the self-energy $\mathbf{\Sigma}^{(n)}$. Note that the dynamical self-energy $\mathbf{\Sigma(\tau)}$ is first constructed on the imaginary time axis, and then transformed to the Matsubara frequency axis using \cref{eq:fourier}.

    \item The new Green's function $\mathbf{G}^{(n+1)}$ is computed by substituting $\mathbf{\Sigma}^{(n)}$ in \cref{eq:dyson}, an equivalent of the Dyson equation.
    The chemical potential is fixed by ensuring the correct particle number in the associated one-particle density matrix,
    \begin{equation}
        \mathbf{\rho}^{(n+1)} = - \mathbf{G}^{(n+1)} (\tau = \beta^{-}).
        \label{eq:build-dm}
    \end{equation}

    \item Calculate the one- and two-body contributions to the total energy,
    \begin{subequations}
        \label{eq:total-energy}
        \begin{align}
            E_\mathrm{Total} &= E_\mathrm{nuclear} + E_{1} + E_{2},
            \\
            E_{1} &= \frac{1}{2} \mathrm{Tr} \left[
                \rho \left(
                    \mathbf{H}_0 + \mathbf{F}
                \right)
            \right],
            \\
            E_{2} &= \sum_{n = -\infty}^{\infty} \mathrm{Tr} \left[
                \mathbf{G} (i\omega_n) \mathbf{\Sigma}(i \omega_n)
            \right].
        \end{align}
    \end{subequations}
    Here, $E_1$ represents the Hartree-Fock contribution, while $E_2$ incorporates dynamical correlation through the Galitskii-Migdal formula~\cite{galitskii_application_1958,holm_total_2000}.
    We consider the self-consistent loop to be converged if the change in both the one-particle and total energy in the current iteration, compared with the previous one, is smaller than specified threshold.
    Otherwise, we repeat steps 2-5.
\end{enumerate}
In all the \scgw calculations presented here, the relative fluctuation in particle number is enforced to be less than $10^{-8}$.
Meanwhile, an absolute threshold of $10^{-6}$ a.u. is used as the convergence criterion of total energies.

\subsubsection{\label{subsec:g0w0}One-shot $GW$}
The premise of one-shot $GW$ is that the reference mean-field calculation serves as a very good starting point. While this assumption may be reasonably accurate for some weakly correlated systems, in general, it introduces a strong dependence of the $\gowo$ results on the mean-field reference.
Following common practice in $\gowo$, we also employ the diagonal approximation here.
Our $\gowo$ calculations can be described with the following steps:
\begin{enumerate}
    \item Starting with the same initial Green's function $G_0$ as in \scgw, we construct the dynamical part of the self-energy $\Sigma$ on the imaginary axis.
    In particular, we are only interested in $\Sigma_{ii} (i \omega_n) = \braket{\psi_i | \Sigma (i \omega_n) | \psi_i}$, where $\ket{\psi_i}$ is the $i$th eigenvector of the mean-field Fock matrix $\mathbf{F}^{(0)}$ with the eigenvalue $\epsilon^0_i$.

    \item The diagonal of the dynamical self-energy on the Matsubara axis is then analytically continued to the real frequency axis.

    \item Using the density $\rho_0$ from the initial mean-field calculation, construct the effective Fock exchange potential $V_x$.

    \item Solve the quasiparticle equation for the $i$th spin-orbital:
    \begin{multline}
        \qquad
        \epsilon^{GW}_i - \epsilon^0_i =
        \\
        \braket{
            \psi_i |
            \left[
                \Sigma_{ii} (\epsilon^{GW}_i - \mu) + V_x - V_{xc}
            \right] |
            \psi_i
        },
        \label{eq:qp-equation}
    \end{multline}
    where $\epsilon^{0}_i$ are the eigenvalues of the underlying mean-field reference, and $V_{xc}$ is the exchange-correlation potential, and $\mu$ is the chemical potential for $G_0$.
\end{enumerate}
In addition to quasiparticle energies, the spectral function is another quantity of interest and provides information about the satellites and peak widths. The $\gowo$ spectral function is defined as
\begin{subequations}
    \label{eq:g0w0-spectral}
    \begin{align}
        A_{G_0W_0} (\omega) &= - \frac{1}{\pi} \textrm{Im} \, \textrm{Tr} \, \mathbf{G}_{G_0W_0} (\omega),
        \\
        \mathbf{G}_{G_0W_0}^{-1} (\omega) &= (\omega + \mu) \mathbf{S} - \mathbf{F}^{(0)} \nonumber
        \\
        & \quad - \mathbf{V}_x - \textrm{diag} \mathbf{\Sigma}(\omega) + \mathbf{V}_{xc}.
        \label{eq:g0w0-diag-sigma-spectral}
    \end{align}
\end{subequations}
For \gowo spectral functions studied here, we observe no noticeable difference between analytical continuation of the full Green's function, and reconstruction of $G(\omega)$ from the diagonal of self-energy, as in \cref{eq:g0w0-diag-sigma-spectral}.
For initializing all the calculations, we use a DFT mean-field reference with the PBE functional for solids, and a HF reference for molecules.

\subsection{\label{subsec:qsgw}Quasiparticle self-consistent \gw-I}
The quasiparticle self-consistent $GW$ (\qsgw) method, proposed by van Schilfgaarde et al. \cite{vanschilfgaarde_quasiparticle_2006,kotani_quasiparticle_2007}, aims to address the issue of reference dependence in $\gowo$, while maintaining accuracy in band gap predictions.
In \qsgw, an effective static exchange-correlation potential $V_{\mathrm{qs}GW}^{\mathrm{eff}}$ is constructed from the dynamical self-energy at each iteration.
Using this effective potential, a quasiparticle equation, similar to \cref{eq:qp-equation}, is solved.
In the original work in Ref.~\onlinecite{vanschilfgaarde_quasiparticle_2006},
two different versions of \qsgw were proposed, labelled as scheme A and B. Here, we limit our discussion to the latter, which is easier to implement within the imaginary-time framework.
The algorithm used in this work closely follows the one employed in Ref.~\onlinecite{forster_low-order_2021}.
We initialize our calculation with the mean-field quasiparticle eigenvalues $\epsilon_i^{(0)}$ and eigenvectors $\mathbf{C}^{(0)}$, along with the corresponding Green's function $\mathbf{G}^{(0)}$ and density matrix $\mathbf{\rho}^{(0)}$ and adopt the following algorithm:
\begin{enumerate}
    \item At the $n$th iteration, construct the dynamical self-energy $\Sigma^{(n)} (i \omega)$ on the Matsubara axis using \cref{eq:hedin-gw-selfenergy}. This self-energy is then transformed from the AO basis to the quasiparticle basis as,
    \begin{equation}
        \mathbf{\Sigma}^{(n)}_{\mathrm{qp}} (i \omega) = \mathbf{C}^{(n-1), \dagger} \mathbf{\Sigma}^{(n)} (i \omega) \mathbf{C}^{(n-1)}.
    \end{equation}

    \item The diagonal of $\Sigma^{(n)}_{ \mathrm{qp}} (i \omega)$ is analytically continued to the real frequency axis using Pad\'{e} or Nevanlinna method.

    \item We then construct the effective exchange-correlation potential using the description:
    \begin{subequations}
        \begin{align}
            \left[ V_{\mathrm{qs}GW}^{\mathrm{eff}, (n)} \right]_{ii}
            &= \left[ \Sigma^{(n)}_{ \mathrm{qp}} \right]_{ii} \left( \omega = \epsilon^{(n-1)}_i \right),
            \\
            \left[ V_{\mathrm{qs}GW}^{\mathrm{eff}, (n)} \right]_{ij}
            &= \left[ \Sigma^{(n)}_{ \mathrm{qp}} \right]_{ij} (\omega = 0) \quad \forall i \neq j.
        \end{align}
    \end{subequations}
    For the diagonal entries, we utilize analytically continued self-energy, whereas in off-diagonal terms arising from $\omega = 0$, we leverage the fact that $\Sigma (\omega = 0) = \Sigma (i \omega = 0)$.
    Therefore, the zero-frequency contribution is obtained by interpolating the Matsubara self-energy using a four-point quadrature.
    This avoids any potential noise that may be introduced in AC.

    \item Similarly, using the density matrix $\rho^{(n-1)}$ from previous iteration, we construct the static self-energy $\Sigma^{ (n)}_{\infty}$, which contains the Hartree and static exchange terms. With this, we define the new Fock operator as:
    \begin{equation}
        \mathbf{F}^{(n)} = \mathbf{H}_0 + \mathbf{\Sigma}^{ (n)}_{\mathbf{\infty}} + \mathbf{V}_{\mathrm{qs}GW}^{\mathrm{eff}, (n)}.
    \end{equation}

    \item The eigenvalues $\mathbf{\epsilon}^{(n)}$ of the effective Fock matrix $\mathbf{F}^{(n)}$ represent the updated quasiparticle energies, while the eigenvectors $\mathbf{C}^{(n)}$ can be used to construct the new density matrix and Green's function as:
    \begin{subequations}
        \begin{align}
            \rho^{(n)}_{\mu \nu}
            &= \sum_{i \in \mathrm{occ}} C^{(n)}_{\mu i} C^{*, (n)}_{\nu i},
            \\
            \mathbf{G}^{(n)} (i \omega)
            &= \left[ (i \omega + \mu) \mathbf{S} - \mathbf{F}^{(n)} \right]^{-1},
        \end{align}
    \end{subequations}
    where the chemical potential $\mu$ is adjusted to enforce the correct number of electrons.
    Note that in this algorithm, the diagonalization of Fock matrix is performed only once per \qsgw iteration.
    Alternatively, as suggested in the original formulation of \qsgw, the quasiparticle equation can be solved iteratively to achieve self-consistency between $F^{(n)}$ and $\rho^{(n)}$.
    However, the absence of this inner self-consistency has been shown not to bear no impact on the final results.~\cite{forster_low-order_2020,forster_low-order_2021}

    \item These steps are repeated until convergence is reached, for which, one may look at the difference in the density matrix,
    \begin{equation}
        \Delta \rho = \frac{1}{N_k} \frac{1}{2 N_{\mathrm{AO}}} \left | \rho^{(n)} - \rho^{(n-1)} \right |.
    \end{equation}
    In practice, however, we find it easier to enforce convergence based on a combination of total energy and band gap (IP) for solids (molecules).
    To enable a better convergence, we also damp the effective \qsgw potential such that at the $n$th iteration, we have
    \begin{equation}
        V_{\mathrm{qs}GW}^{\mathrm{eff}} = \alpha V_{\mathrm{qs}GW}^{\mathrm{eff}, (n)} + (1 - \alpha) V_{\mathrm{qs}GW}^{\mathrm{eff}, (n-1)},
    \end{equation}
    where $\alpha = 0.7$ suffices in most cases.
\end{enumerate}
The converged quasiparticle energies can be used to calculate the band gap, IP, and density of states (or spectral function).
We would like to note that while tight convergence thresholds are desirable, in practice, for solids we consider a \qsgw calculation  converged if the difference in successive band gaps is less than 0.01 eV and that in total energy is less than 1 $mE_h$.
For molecules, the problems are less severe and converging IPs to within 1 meV is generally possible.

\subsection{Quasiparticle self-consistent \gw-II}
The \qpgw approximation, also referred as the QP-2 self-consistency scheme in the literature, reduces the dynamical self-energy to an effective static term by a first-order Taylor-series expansion.
Since the scheme was originally proposed in the Matsubara formalism, our implementation avoids the additional dependence on analytic continuation.
Like all other methods, \qpgw is also initialized with mean-field Fock and density matrices, $\mathbf{F}^{(0)}$ and $\rho^{(0)}$.
The iterations of \qpgw can be described as:
\begin{enumerate}
    \item Construct the static and dynamical parts of the self-energy $\Sigma_c^{(n)}$ for $n$th iteration, and update the Fock matrix as
    \begin{equation}
        \mathbf{F}^{(n)} = \mathbf{H}_0 + \mathbf{\Sigma}^{(n)}_{\infty}.
    \end{equation}
    This step is similar to the first step of \scgw algorithm.

    \item Diagonalize $\mathbf{F}^{(n)}$ to obtain molecular orbitals $\mathbf{C}^{(n)}$ and eigenvalues $\mathbf{\epsilon}^{(n)}$.

    \item The first non-trivial step in \qpgw is the linear-order Taylor expansion of the Matsubara self-energy:
    \begin{equation}
        \Sigma(i \omega) \simeq
        \Sigma_0
        + \left. \frac{
            \partial \Sigma(i\omega)
        }{
            \partial (i\omega)
        } \right \vert_{\omega = 0} (i \omega),
    \end{equation}
    where $\Sigma_0 = \Sigma(i \omega = 0)$.
    Substituting this approximation in \cref{eq:hedin-gw-dyson}, we get
    \begin{equation}
        \mathbf{G}^{-1} (i\omega)
        =
        i \omega \mathbf{Z}^{-1} + \mu \mathbf{S} - \mathbf{F} - \mathbf{\Sigma}_0,
        \label{eq:dyson-ao-basis}
    \end{equation}
    where the quasiparticle renormalization (or mass) term $\mathbf{Z}^{-1}$ is defined as
    \begin{equation}
        \mathbf{Z}^{-1}
        =
        \mathbf{S} - \left. 
            \frac{\partial \mathbf{\Sigma}}{\partial (i\omega)}
        \right \vert_{\omega = 0}.
    \end{equation}

    \item In the orthonormal basis composed of molecular orbitals $\mathbf{C}^{(n)}$ (obtained in step 2),  
    the Dyson equation reads as
    \begin{equation}
        \mathbf{G}^{-1}_\textrm{MO} (i\omega)
        =
        i \omega \mathbf{Z}^{-1}_\textrm{MO} + \mu - \mathbf{\epsilon} - \mathbf{\Sigma}_{0, \textrm{MO}}.
        \label{eq:dyson-mo-basis}
    \end{equation}
    The transformation of every quantity $\mathbf{X}$ in \cref{eq:dyson-ao-basis} to $\mathbf{X}_\textrm{MO}$ in \cref{eq:dyson-mo-basis} is defined as
    \begin{equation}
        \mathbf{X}_\textrm{MO} = \mathbf{C}^{(n) \dagger} \mathbf{X} \mathbf{C}^{(n)}.
    \end{equation}
    Readjusting the terms in \cref{eq:dyson-mo-basis}, we obtain:
    \begin{equation}
        \mathbf{Z}^{1/2}_\textrm{MO} \mathbf{G}^{-1}_\textrm{MO} \mathbf{Z}^{1/2}_\textrm{MO}
        =
        i \omega - \mathbf{H}_{\textrm{eff,MO}},
        \label{eq:qpgw-intermediate}
    \end{equation}
    where the effective Hamiltonian is defined as:
    \begin{equation}
        \mathbf{H}_\textrm{eff,MO} = \mathbf{Z}^{1/2}_\textrm{MO} \left[
            \mathbf{\epsilon} - \mu + \mathbf{\Sigma}_{0, \textrm{MO}}
        \right] \mathbf{Z}^{1/2}_\textrm{MO}.
        \label{eq:qpgw-hamiltonian}
    \end{equation}

    \item The second, and perhaps more serious, approximation invoked in \qpgw is $\mathbf{Z}_\textrm{MO}^{-1} \approx \mathbb{I}$, where $\mathbb{I}$ is the identity matrix. Consequently, we drop the quasiparticle mass from the left hand side of \cref{eq:qpgw-intermediate}.
    However, the dependence of the effective Hamiltonian on $\mathbf{Z}_\textrm{MO}^{-1}$ in \cref{eq:qpgw-hamiltonian} is retained.
    Finally, we obtain a new Green's function with an approximated Dyson equation as
    \begin{equation}
        \left[\mathbf{G}^{(n+1)}\right]^{-1} (i \omega)
        =
        i \omega \mathbf{S} - \mathbf{H}_{\textrm{eff}}^{(n)},
        \label{eq:qpgw-dyson}
    \end{equation}
    with the AO-basis effective Hamiltonian defined as
    \begin{equation}
        \mathbf{H}_{\textrm{eff}}^{(n)} = \mathbf{S} \mathbf{\zeta}^{(n)} \left(
            \mathbf{F}^{(n)} - \mu \mathbf{S} + \mathbf{\Sigma}^{(n)}_0
        \right) \mathbf{\zeta}^{(n)} \mathbf{S}.
    \end{equation}
    Here, we have as an intermediate
    \begin{equation}
        \mathbf{\zeta}^{(n)} = \mathbf{C}^{(n)} \left[\mathbf{Z}^{(n)}_\textrm{MO}\right]^{1/2} \mathbf{C}^{(n)\dagger}.
    \end{equation}

    \item Once again, the chemical potential $\mu$ is determined to maintain the correct number of electrons, and the algorithm is repeated until numerical self-consistency is achieved in the total energy. Furthermore, due to the absence of the dynamical self-energy, the Galitski-Migdal contribution $E_2$ vanishes in \cref{eq:total-energy}.
\end{enumerate}
The outcome of a converged \qpgw calculation is the effective Hamiltonian whose eigenvalue correspond to the quasiparticle energies.
As a result, no analytical continuation is required in this approach.
Finally, we note that convergence thresholds similar to those in \scgw are enforced for \qpgw as well.

\section{\label{sec:comp-details}Computational details}
The primary goal of our work is to compare different versions of the $GW$ method, while keeping implementation details fairly uniform.
To ensure that, we employ Gaussian orbitals (or their periodic extension) as a single-particle basis of atomic orbitals.
In particular, for solids, we use the \texttt{GTH-DZVP-MOLOPT-SR} basis-sets, with the corresponding \texttt{GTH-PBE} pseudopotentials.
For molecular test-set, we perform all-electron calculations using the \texttt{cc-pVQZ} basis-sets.
Basis-set extrapolation is not performed for either set of systems.
For two-electron integrals, we use the density-fitting approximation~\cite{stoychev_automatic_2017} implemented in PySCF package (version 2.0.1)~\cite{sun_pyscf_2018, sun_recent_2020}.
Specifically, we use the \texttt{def2-tzvp-jkfit} and \texttt{cc-pVQZ-jkfit} basis-sets as the auxiliary basis-set for density-fitting in solids and molecules, respectively.
We note that in \texttt{GTH} pseudopotentials for Zn and Ga, the valence shell $3d$ orbital is considered as a part of the core. As a result, the correlation effects arising from these orbitals are not well recovered.

Dynamical quantities that depend on the imaginary time and frequency are represented using highly accurate yet compact intermediate-representation (IR) grids~\cite{li_sparse_2020,shinaoka_efficient_2021}.
All the $GW$ calculations are performed at an inverse temperature $\beta = 1000 \, \textrm{a.u.}^{-1}$.
For the analytical continuation of Matsubara Green's function and self-energy diagonal, we use both Nevanlinna~\cite{fei_nevanlinna_2021} and Pad\'{e}~\cite{vidberg_solving_1977} methods, often interchangeably, and find that they provide almost identical results for quasiparticle peaks.
Using the Pad\'{e} approach also allows us to establish equivalence with previous implementations.
In order to avoid poles on the real axis, when performing analytical continuation, we keep a small imaginary component $\eta$ for the frequency, i.e.,
\begin{equation}
    X (i \omega_n) \xrightarrow[\textrm{Continuation}]{\textrm{Analytical}} X (\omega + i \eta).
\end{equation}
This leads to a broadening of the spectral features which is inherently different from the broadening induced by the imaginary part of self-energy.
For obtaining band gaps from \scgw, we employ $\eta = 0.001$ a.u., which provides sharp and well-resolved quasiparticle peaks.
For the AC step in \qsgw, for most systems, we use $\eta = 0.01$ a.u.

Other than Ewald summation for the Fock exchange~\cite{paier_perdewburkeernzerhof_2005,broqvist_hybrid-functional_2009,sundararaman_regularization_2013}, no finite-size corrections have been used for any of the $GW$ or DFT results.
Instead, band gaps have been extrapolated from calculations with $4\times 4\times 4$ and $6\times 6 \times 6$ $k$-point samplings in the Brillouin zone (BZ), assuming a linear relation between the band gap and the inverse cube root of the number of $k$-points in the BZ.
For carbon and silicon, calculations with $6 \times 6 \times 6$ and $8 \times 8 \times 8$ $k$-points were used for extrapolation.
We also note that all DFT band gaps were calculated using a $6\times 6\times 6$ $k$-points sampling.

Before we proceed to the Results section, it is worthwhile to comment on the need for accurate grids for dynamical quantities, especially in a \scgw calculation.
In a typical cycle of \gw, one needs to transform $G$, $\Sigma$ and $\Pi$, between the imaginary time and Matsubara frequency axes (cf. \cref{subsubsec:scgw}).
Furthermore, adjusting the chemical potential triggers a fair amount of $G(i\omega) \leftrightarrow G(\tau)$ operations.
The Green's function would quickly accumulate detrimental amount of noise if the underlying grids do not guarantee a high accuracy in Fourier transforms.
In this work, most of the calculations employ 104 IR grid points, which ensures an accuracy of $10^{-10}$ or better in Fourier transformations.
This helps us enforce tight thresholds for total energy and particle number in \scgw, necessary for reliable applications to real systems.
Older generation of \gw codes usually employ a uniform Matsubara grid, which requires several hundreds or up to a thousand grid points to ensure high accuracy.
An example is the ComDMFT~\cite{choi_comdmft_2019} code used in Kutepov's comparative study~\cite{kutepov_self-consistent_2017}.
Similarly, for Grumet's results in Ref.~\onlinecite{grumet_beyond_2018}, the early generation of sparse grids~\cite{kaltak_low_2014} appears to afford merely 5-6 iterations of \scgw and a particle number fluctuation of $\sim 10^{-3}$.
Development of accurate imaginary time and Matsubara grids has been an active area of research~\cite{kaltak_low_2014,gull_chebyshev_2018, hugo_grid,li_sparse_2020,kaye_discrete_2022,sheng_low-rank_2023}, culminating in the developments of accurate grids such as IR~\cite{li_sparse_2020} and discrete Lehmann representation~\cite{kaye_discrete_2022} grids.

\begin{figure}[tbp]
    \centering
    \includegraphics[width=\linewidth]{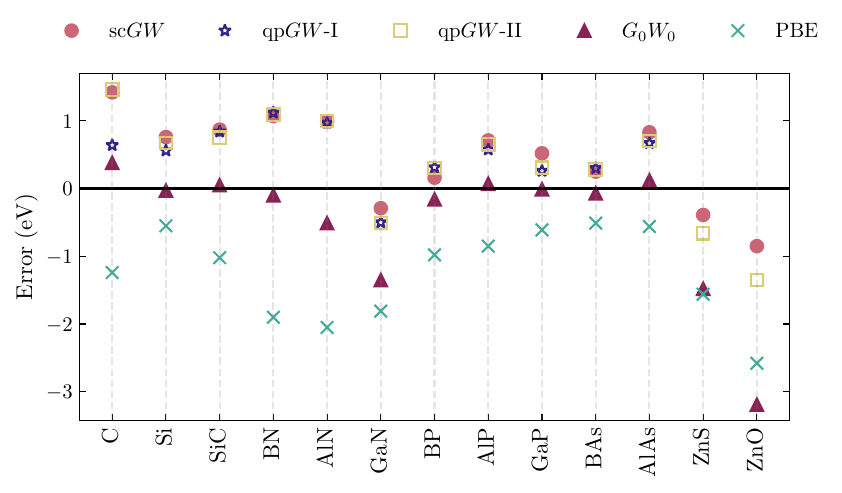}
    \caption{Band gap errors, with respect to experiment, for selected semiconductors and insulators, as predicted by full and quasiparticle self-consistent \gw methods. PBE and \gowo-PBE results are also included for reference.}
    \label{fig:band_gap_error}
\end{figure}

\begin{table*}[htbp]
    \caption{Band gaps (in eV) for selected semiconductors and insulators, comparing PBE, \gowo-PBE, quasiparticle and fully self-consistent \gw methods against experimental data. Previously reported values for \scgw, \qsgw, and \qpgw are also included for reference. For ZnO and ZnS, the \qsgw calculation could not be converged. We have used diamond lattice structure for C and Si, and zincblende for all other systems. Lattice constants $a$ are also reported here. Experimental values for band gaps have been adopted from Refs.~\onlinecite{LandoltBornstein1989,thompson_deposition_2001,Roppischer2009,grumet_beyond_2018,Kang2019}.}
    \label{tab:band_gaps}
    \centering

    \settowidth{\templen}{\gowo-PBE}

    \begin{tabular}{l|P{1cm}|P{1cm}|P{1cm}|c|c|P{1cm}|c|c|c|c|c|c|P{1cm}}
        \hline \hline
        \multirow{2}{*}{Material} & \multirow{2}{*}{PBE} & \multirow{2}{*}{$\gowo$} & \multirow{2}{*}{\scgw} & \multirow{2}{*}{\qsgw} & 
            \multirow{2}{*}{\qpgw} & \multirow{2}{*}{Exp.} & \multicolumn{2}{c|}{Other \scgw} & \multicolumn{4}{c|}{Other qp$GW$-I/II} & \multirow{2}{*}{$a$ ($\mathrm{\AA}$)}   \\
        \cline{8-13}
        & & & & & & & Ref.~\onlinecite{grumet_beyond_2018} & Ref.~\onlinecite{kutepov_self-consistent_2017} & Ref.~\onlinecite{grumet_beyond_2018} & Ref.~\onlinecite{kutepov_self-consistent_2017} & Ref.~\onlinecite{kotani_quasiparticle_2007} & Ref.~\onlinecite{lei_gaussian-based_2022} & \\
        \hline
        C       & 4.24  & 5.86  & 6.90  & 6.12  & 6.94  & 5.48          & 6.41  & 6.15  & 6.43  & 6.18  & 5.97  & 6.14  & 3.56  \\
        Si      & 0.62  & 1.14  & 1.93  & 1.73  & 1.84  & 1.17          & 2.18  & 1.55  & 1.49  & 1.41  & 1.28  & 1.32  & 5.43  \\
        SiC     & 1.38  & 2.45  & 3.27  & 3.24  & 3.15  & 2.40          & 3.29  & 2.89  & 2.88  & 2.79  & 2.58  & 2.81  & 4.35  \\
        BN      & 4.50  & 6.30  & 7.47  & 7.52  & 7.49  & 6.40          & 7.67  & 7.06  & 7.50  & 7.06  &       & 7.07  & 3.61  \\
        AlN     & 3.29  & 4.83  & 6.33  & 6.31  & 6.34  & 5.34          &       &       &       &       &       &       & 4.37  \\
        GaN     & 1.59  & 2.05  & 3.11  & 2.90  & 2.89  & 3.40          & 3.94  &       & 3.78  &       &       &       & 4.53  \\
        BP      & 1.42  & 2.24  & 2.56  & 2.71  & 2.70  & 2.40          &       &       &       &       &       &       & 4.53  \\
        AlP     & 1.60  & 2.52  & 3.16  & 3.03  & 3.10  & 2.45          & 3.20  & 2.84  & 2.94  & 2.80  &       & 2.76  & 5.46  \\
        GaP     & 1.65  & 2.25  & 2.78  & 2.52  & 2.57  & 2.26          & 2.77  &       & 2.67  &       &       &       & 5.45  \\
        BAs     & 1.31  & 1.75  & 2.07  & 2.12  & 2.10  & 1.82          &       &       &       &       &       &       & 4.77  \\
        AlAs    & 1.60  & 2.28  & 2.99  & 2.83  & 2.86  & 2.16          & 2.98  &       & 2.84  &       &       &       & 5.66  \\
        ZnS     & 2.04  & 2.12  & 3.21  &       & 2.94  & 3.60          & 4.68  & 4.28  & 4.27  & 4.19  & 4.13  &       & 5.42  \\
        ZnO     & 0.66  & 0.05  & 2.39  &       & 1.89  & 3.24          & 4.92  &       & 4.29  &       &       &       & 4.58  \\
        \hline \hline
    \end{tabular}
\end{table*}

\section{\label{sec:results}Results}
Electronic structure methods are best compared when multiple properties in different kinds of systems are analyzed.
For a thorough analysis of the \gw methods, we compare band gaps in insulating solids, as well as IPs in molecules.
Additionally, we take a closer look at the spectral function from different methods, while also scrutinizing their convergence trends.

\subsection{Band gaps in insulators}
First, we analyze the band gaps in selected insulators and semiconductors.
This is captured in \cref{fig:band_gap_error} where we compare \scgw, \qsgw, and \qpgw.
We also show results for PBE and $\gowo$-PBE.
The corresponding numerical data, along with existing reference values from the literature, are also presented in \cref{tab:band_gaps}, along with the crystal structures and lattice constants.
Our results unequivocally show that for band gaps, quasiparticle and fully self-consistent $GW$ methods produce very similar results.
On the other hand, consistent with existing literature, we find that while PBE generally underestimates the band gaps, $\gowo$-PBE appears to be the most accurate among all the methods.
It is sometimes presumed that \scgw severely over-estimates band gaps, while qp$GW$ improves upon it.
However, the results in \cref{tab:band_gaps} do not support such claims.

For ZnS, ZnO and GaN, the behavior is somewhat different, i.e., \gowo does not improve upon PBE, and \scgw appears to perform the best.
In systems with Zn and Ga, the closed-shell $3d$ orbitals, which play an important role in electron correlation, are not present in the \texttt{GTH-DZVP-MOLOPT-SR} basis-set, and instead treated as part of the pseudopotential core.

It is worth mentioning that converging the \gw calculations with respect to basis-set size, in an all-electron setting, generally leads to more accurate band gaps.
For instance, when comparing \scgw band gaps in this work against those in Ref.~\onlinecite{yeh_fully_2022}, it is evident that going from \texttt{GTH} basis-set and pseudopotential to a sufficiently large all-electron basis-set \texttt{x2c-TZVPall} improves the band gaps by $\sim 0.3$ eV.
Nevertheless, we do not expect any significant variation in the relative accuracy between different self-consistent \gw methods.
This is because the \texttt{GTH} basis-sets and pseudopotentials are well optimized (for DFT) with respect to converged plane wave calculations~\cite{goedecker_separable_1996,hartwigsen_relativistic_1998}.
One can expect reasonable behavior for correlated methods as well, particularly since the materials here involve light elements with negligible relativistic effects~\cite{harsha_challenges_2024}.
The overall quality of the different self-consistent \gw results can be expected to get better with larger all-electron basis-sets.

\begin{figure}[hbp]
    \centering
    \includegraphics[width=0.95\linewidth]{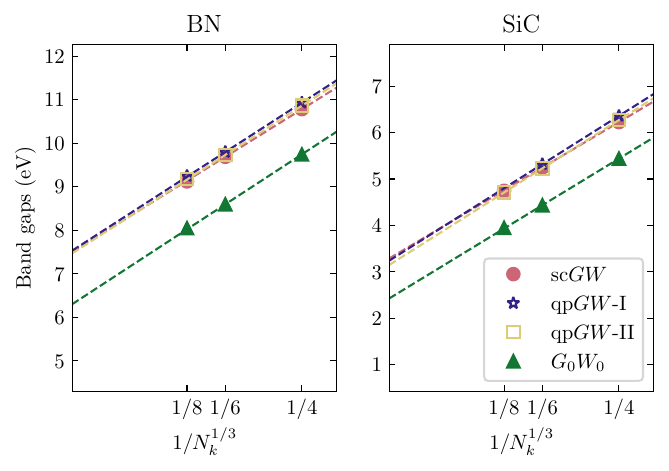}
    \caption{Variation in BN and SiC band gaps with respect to Brillouin-zone sampling. The linear fit is performed using $N_k^{1/3} = 4$ and $6$ grids.
    The results from $N_k^{1/3} = 8$ show excellent agreement with this fit, demonstrating the reliability of the extrapolation procedure.}
    \label{fig:kpt_converge}
\end{figure}

\subsubsection{Finite-size extrapolation}
The band gaps presented here are calculated using linear extrapolation of results from $N_k^{1/3} = 4, 6$ as mentioned in \cref{sec:comp-details}.
To validate that these two grid-samplings are sufficient to extrapolate from, for SiC and BN as examples, we have also performed calculations using an $8 \times 8 \times 8$ $k$-mesh.
The resultant $N_k^{1/3} = 8$ band gaps show excellent agreement with the linear fits derived from $N_k^{1/3} = 4$ and $N_k^{1/3} = 6$ data.
This result is summarized in \cref{fig:kpt_converge}.
The use of linear extrapolation technique ensures that finite-size corrections and associated approximations do not influence the comparison between different methods.

\subsubsection{Convergence trends}
\begin{figure*}[tpbh]
    \centering
    \includegraphics[width=0.9\linewidth]{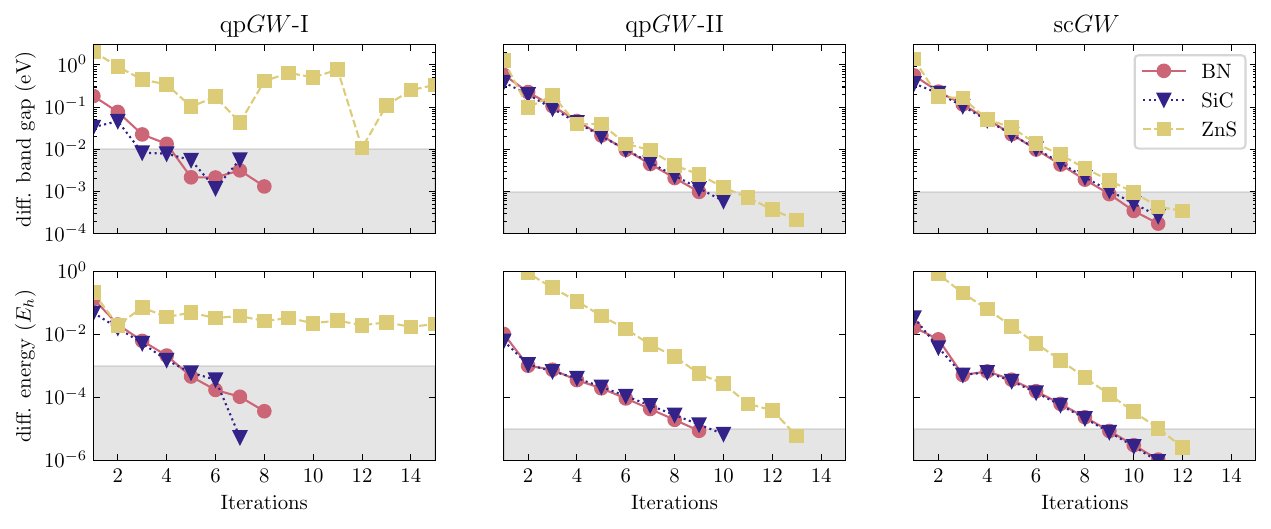}
    \caption{Convergence trends for band gaps (eV) and total energy per unit cell ($E_h$) for \qsgw, \qpgw and \scgw, all calculated for $4\times 4 \times 4$ $k$-mesh sampling in the BZ. We consider SiC and BN as typical examples of easy convergence, while ZnS proves to be a challenging system for \qsgw.} 
    \label{fig:qsgw-conv}
\end{figure*}

In \cref{fig:qsgw-conv}, we study the trends in band gaps and total energy per unit cell for SiC, BN and ZnS as we iterate through different self-consistency schemes.
First, we observe that for all three materials, it is fairly straightforward to impose tight convergence criterion in \scgw and \qpgw.
Even without the use of convergence acceleration techniques such as direct inversion of iterative subspace~\cite{pokhilko_iterative_2022}, both these methods converge the total energy to within $10^{-5}$ a.u., and band gaps to within $10^{-4}$ eV in 10-20 iterations.
Contrarily, extra care is required in \qsgw, where we rely on trends in both band gap (or IPs) and total energy to establish convergence.
For solids, tight convergence was achievable only for small systems like SiC and BN, and for the general case, the band gaps are converged only at the level of 0.01 eV.
For ZnS and ZnO (not shown here), we could not converge \qsgw.
This is likely due to the complexity of electron correlation brought by the transition metal.
Similar challenges have been noted in the literature~\cite{lei_gaussian-based_2022}.

\subsubsection{Spectral functions}
\begin{figure}[b]
    \centering
    \includegraphics[width=0.9\linewidth]{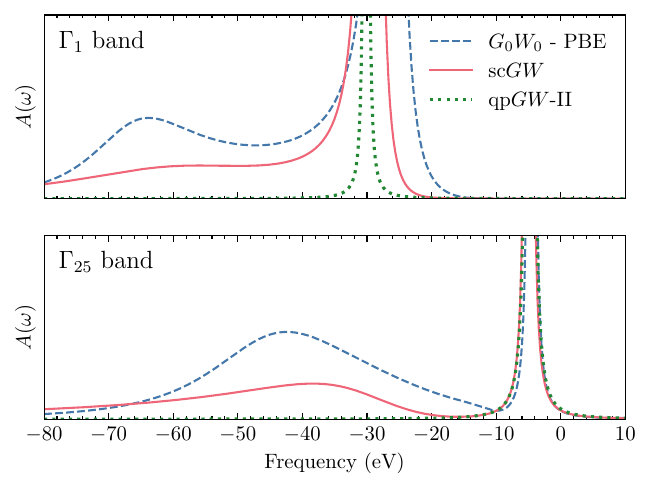}
    \caption{Spectral function for the $\Gamma_1$ and $\Gamma_{25}$ bands in diamond, calculated using \gowo-PBE, \scgw and \qpgw. The results use a $6\times 6\times 6$ $k$-mesh, and Pad\'e analytical continuation for \gowo and \scgw. The $y$-axis is magnified to emphasize the satellites.}
    \label{fig:satellite}
\end{figure}

One of the concerns surrounding \scgw is its inability to describe plasmon satellites, which are, however, captured by $\gowo$.
This behavior was first noted by Holm et al. for the electron gas~\cite{holm_fully_1998}, where iterations of self-consistency were observed to completely diminish sharp satellites. 
Similar trends have been reported for diamond by Grumet et al. \cite{grumet_beyond_2018}.
In \cref{fig:satellite}, we compare $\gowo$-PBE, \scgw and \qpgw spectral functions for diamond for the $\Gamma_1$ and $\Gamma_{25}$ bands, with the first two computed using the Pad\'e continuation.
Strikingly, we find that \scgw is able to capture the plasmon satellites for both the bands. However, the weight of the satellites is smaller than \gowo, supporting existing knowledge that self-consistency leads to an increased quasiparticle weight.
Comparing with similar results in existing literature, we note that our \scgw is capable of predicting satellite features where earlier works could not. We speculate that this improvement is a result of superior quality of time and frequency grids, as well as strong convergence criteria on both particle number and total energies.
These results should be contrasted with quasiparticle self-consistent $GW$ schemes, which, by definition, do not retain any information about satellite structures in the spectral function.

\subsection{Ionization potentials in molecules}
Comparison of different self-consistency schemes such as \scgw and \qsgw have been relatively scarce in the context of molecular systems as well~\cite{koval_fully_2014,caruso_benchmark_2016}.
Moreover, to our knowledge, application of the \qpgw scheme to molecules has not been explored in the literature.
Here, we bridge this gap with our implementation and benchmark various \gw self-consistency schemes for predicting IPs across a selection of 29 molecules.
This molecular dataset was originally used by Maggio et al.~\cite{maggio_gw_2017} to benchmark vertex corrections on \gowo methods.
Recently, Wen et al.~\cite{wen_comparing_2024} also used this dataset to show that \scgw, with the same implementation as is used in this work, provides accurate IPs, comparable in quality with vertex-corrected \gowo methods.

In \Cref{tab:molecules_summary}, we compare the mean absolute error (MAE) and standard deviation of error (SD) with respect to both the experiment and $\Delta$CCSD(T), for the prediction of IP in these molecules.
Complete set of IP values are presented in \cref{tab:molecules_full_data} of \cref{app2:molecule_ips}.
The \gowo-HF, \scgw, and the $\Delta$CCSD(T) results are taken from Ref.~\onlinecite{wen_comparing_2024}, while calculations for the quasiparticle self-consistent methods are carried out with the implementations discussed in Section~\ref{sec:theory}.
Unlike in solids, where full and quasiparticle self-consistency results in similar band gaps, for molecular IPs, \scgw clearly outperforms \qsgw and \qpgw.
With respect to experiment, the MAE of 0.30 eV in \scgw is much smaller than 0.42 eV and 0.44 eV in \qsgw and \qpgw, respectively.
The standard deviations in all the methods, including \gowo-HF, are similar in magnitude.
It is interesting to note that with respect to $\Delta$CCSD(T), the quasiparticle and fully self-consistent $GW$ methods exhibit comparable MAE and SD.
We also note that the \qsgw and \qpgw results are very close to each other for molecular IPs.
As mentioned in \cref{sec:comp-details}, all the results have been calculated using the \texttt{cc-pVQZ} basis.
Considering the large size of this basis, we anticipate that extrapolation to complete basis-set limit will not introduce any significant change in overall trends.

\begin{table}[tbp]
    \caption{Mean absolute error (MAE) and associated standard deviation (SD) in eV for different self-consistent $GW$ methods for the 29-molecule data-set. $\Delta$CCSD(T) results are also included for reference. $^a$Results adopted from Ref.~\onlinecite{wen_comparing_2024}.}
    \label{tab:molecules_summary}
    \centering

    \settowidth{\templen}{w.r.t. $\Delta$CCSD(T)}
    \setlength{\templen}{\dimexpr(\templen-2\tabcolsep)/2}

    \begin{tabular}{l|  *{2}{>{\centering\arraybackslash}p{\templen}} | *{2}{>{\centering\arraybackslash}p{\templen}}}
        \hline \hline
        \multirow{2}{*}{Method} & \multicolumn{2}{c|}{w.r.t experiment} & \multicolumn{2}{c}{w.r.t. $\Delta$CCSD(T)} \\
        \cline{2-5}
        & MAE & SD & MAE & SD \\
        \hline
        \gowo-HF$^a$ & 0.65 & 0.36 & 0.49 & 0.28 \\
        sc$GW^a$ & 0.30 & 0.27 & 0.24 & 0.20 \\
        \qsgw & 0.42 & 0.32 & 0.28 & 0.14 \\
        \qpgw & 0.44 & 0.31 & 0.29 & 0.16 \\
        $\Delta$CCSD(T)$^a$ & 0.23 & 0.32 & - & - \\
        \hline \hline
    \end{tabular}
\end{table}

In terms of convergence, \qsgw in molecules exhibits a much better behavior, and in most cases, IPs can be systematically converged to less than $10^{-3}$ eV in 10-20 iterations.
With the exception of CO, SiO, and CS, all other IPs are well converged.
For \scgw and \qpgw, we do not encounter any convergence issues.

\section{Conclusion}
We have presented a comprehensive analysis of quasiparticle self-consistent \gw methods against their one-shot and fully self-consistent counterparts.
We have employed finite-temperature Green's functions and maintained identical technical details for all the methods, ensuring a fair comparison free of inconsistencies arising from differences in implementations.
Our aim is to compare different self-consistent schemes available in the GW community and our observations are based on analysis of three key measures, namely (i) band gap recovery for solids and IP recovery for molecules, (ii) convergence characteristics of each of the schemes, and (iii) recovery of spectral features.

The band gap recovery, for simple semiconductors considered in a Gaussian basis, was very similar for all the self-consistency schemes.
These observations reinforce findings in Refs.~\onlinecite{grumet_beyond_2018,kutepov_self-consistent_2017} who also observed that the band gaps from \qsgw and \qpgw schemes are very close to \scgw values.
When molecules are considered, we found that \scgw was clearly superior to \qsgw and \qpgw for the recovery of the IPs.
Furthermore, both the qp\gw methods produced results of similar quality.

We have observed that while all the values of the band gaps remained relatively agnostic to the self-consistency scheme, the convergence characteristics did not. 
For solids, we experienced much more difficulties converging the \qsgw scheme than \scgw and 
\qpgw. Moreover, we were unable to converge \qsgw to the same accuracy as the two other schemes. 
For molecules, we observed fewer convergence difficulties in the \qsgw scheme and it was generally possible to converge IPs to less than $10^{-3}$ eV.

The primary quasiparticle peaks are well recovered in all the self-consistency schemes.
Both \qsgw and \qpgw are by definition incapable of illustrating any other features than these primary peaks in the spectrum.
Remarkably, the plasmon satellites are preserved by \scgw in the diamond lattice carbon studied here, although their weight is reduced in comparison with \gowo results.
Comparing with earlier works, such as Ref.~\onlinecite{grumet_beyond_2018}, we speculate that this improvement in our version of \scgw is a result of superior quality of time and frequency grids, as well as strong convergence criteria on both particle number and total energies which were not present in the earlier works. 
It should be noted however that despite this success of \scgw in the recovery of the plasmon satellites, we can still expect that accurate illustration of some of the spectral features will require higher-level Feynman diagrams than the ones present in the \gw self-energy.

Furthermore, the evaluation of total energy makes sense only for \scgw, where thermodynamic properties are uniquely defined.
The absence of reliable total energy may not be a concern in solids since the cohesive energy of a unit cell is rarely of interest.
In contrast, the ability to calculate energies is of paramount importance in molecules since most calculations rely on the evaluation of energy differences.
However, even when the calculation of total energy or energy differences may be avoided, other observable quantities, even as simple as orbital occupation or more complicated correlation functions are also uniquely defined only for \scgw.

Considering all these observations and knowing that the development of highly accurate, yet compact, grids together with additional integral factorization schemes make the \scgw scheme already quite computationally accessible, one can arrive at following practical recommendations on when the different schemes should be used. 
Only in lighter accuracy calculations when the tight convergence of the IP or band gap values is not necessary, and only primary quasiparticle peaks are of interest, and additionally only very limited in memory is available, \qsgw or \qpgw may be beneficial since in this method at the start of each iteration, the Green's function is reset to a mean-field object, thus partially alleviating some of the memory bottlenecks. 
In all the other cases, \scgw method will be superior due to smoother convergence, good band gap recovery, possibility of observing more spectral features, and well defined evaluation of thermodynamic quantities including energy.

\begin{acknowledgments}
    This material is based upon work supported by the U.S. Department of Energy, Office of Science, Office of Advanced Scientific Computing Research and Office of Basic Energy Sciences, Scientific Discovery through Advanced Computing (SciDAC) program under Award Number DE-SC0022198.
    This research used resources of the National Energy Research Scientific Computing Center, a DOE Office of Science User Facility supported by the Office of Science of the U.S. Department of Energy under Contract No. DE-AC02-05CH11231 using NERSC award BES-ERCAP0029462 and BES-ERCAP0024293.
    We also acknowledge helpful discussion with Dr. Chia-Nan Yeh and Dr. Arno F\"{o}rster.
\end{acknowledgments}

\appendix

\section{\label{app2:molecule_ips}Ionization potential data for molecules}
Detailed ionization potential data for the 29 molecules studied in \cref{tab:molecules_summary} is presented here in \cref{tab:molecules_full_data}.
For N$_2$, ClF and HClO, the \qsgw loop (marked with a * in \cref{tab:molecules_full_data}) was terminated once the IP was converged to within 0.001 eV.
For CO, SiO and CS, we could not converge the \qsgw results (marked with ** in \cref{tab:molecules_full_data}) even after 20 iterations.
For all these calculations, we have employed the \texttt{cc-pVQZ} basis-set.
Furthermore, all calculated IPs assume vertical ionization, i.e., the geometry of the ionized molecule is not relaxed.
Among experimental results, the vertical IP values are shown in italics.
The discrepancy between experiment and theory is generally larger when the IPs that are non-vertical in nature.
For such systems, comparing the \gw methods against $\Delta$CCSD(T), while assuming vertical ionization, may be more sensible.

\begin{table*}[htbp]
    \centering
    \caption{The first ionization potential (eV) data for 29 molecules. All the calculations were performed using \texttt{cc-pVQZ} basis-set. Italicized values are for the vertical IPs. $^*$Only converged in IP values, not total energies. $^{**}$Not converged.}
    \label{tab:molecules_full_data}
    \begin{tabular}{l|c|c|c|c|c|c}
        \hline \hline
        Molecule    & $\Delta$CCSD(T)~\cite{maggio_gw_2017}   & \gowo-HF~\cite{wen_comparing_2024}   & \scgw\cite{wen_comparing_2024}     & \qsgw      &\qpgw     & Exp.~\cite{mccormackMeasurementHighRydberg1989,dugourdMeasurementsLithiumCluster1992,tricklStateSelectiveIonization1989,k.bulginHeIPhotoelectronSpectrum1976,dykeIonizationEnergiesDiatomic1984,bieri304nmHe1980,robergeFarUltravioletHeI1978,liasIonEnergeticsData,baumgaertelPhotoelectronSpectraMolecular1989,cowleyLewisBaseBehavior1982,bieri304nmHe1982,bannaMolecularPhotoelectronSpectroscopy1976,pottsPhotoelectronStudiesIonic1977,kreileExperimentalTheoreticalInvestigation1982,vovnaPhotoelectronSpectraHydrazine1975,vorobevMassSpectrometryReasonance1989,ashmoreStudyMediumSize1977,kimuraHandbookHePhotoelectron1981,elandPhotoionizationMassSpectrometry1977,pottsObservationForbiddenTransitions1974,cradockPhotoelectronSpectraMethyl1972,nakasgawaMassSpectrometricStudy1981,kingPhotoelectronSpectrumShortlived1972,colbournePhotoelectronSpectraIsoelectronic1978}      \\
        \hline
        H$_2$           & -16.39 & -16.59 & -16.24 & -16.53         &-16.55     & -15.43 \\
        Li$_2$          & -5.17  & -5.38  & -4.94  & -5.37          &-5.34      & -4.73  \\
        N$_2$           & -15.49 & -16.56 & -15.57 & -15.98$^*$   &-16.06     & -15.58 \\
        P$_2$           & -10.76 & -10.71 & -9.85  & -10.60         &-10.51     & -10.62 \\
        Cl$_2$          & -11.62 & -11.98 & -11.24 & -11.84         &-11.72     & \textit{-11.49} \\
        CH$_4$          & -14.40 & -14.92 & -14.32 & -14.68         &-14.76     & \textit{-13.6}  \\
        C$_2$H$_4$      & -10.69 & -10.88 & -10.18 & -10.65         &-10.66     & \textit{-10.68} \\
        C$_2$H$_2$      & -11.42 & -11.77 & -10.96 & -11.50         &-11.54     & \textit{-11.49} \\
        SiH$_4$         & -12.82 & -13.35 & -12.80 & -13.17         &-13.20     & \textit{-12.3}  \\
        LiH             & -7.94  & -8.28  & -7.94  & -8.31          &-8.22      & -7.9   \\
        NH$_3$          & -10.92 & -11.40 & -10.86 & -11.17         &-11.23     & $-10.82$ \\
        PH$_3$          & -10.49 & -10.92 & -10.28 & -10.75         &-10.75     & \textit{-10.59} \\
        H$_2$S          & -10.43 & -10.69 & -10.09 & -10.51         &-10.55     & \textit{-10.50}  \\
        HF              & -16.09 & -16.49 & -16.26 & -16.52         &-16.59     & \textit{-16.12} \\
        NaCl            & -9.13  & -9.43  & -8.93  & -9.48          &-9.33      & \textit{-9.80}   \\
        HCN             & -13.64 & -14.07 & -13.22 & -13.81         &-13.82     & \textit{-13.61} \\
        N$_2$H$_4$      & -10.24 & -10.38 & -9.71  & -10.11         &-10.16     & \textit{-8.98}  \\
        CH$_3$OH        & -11.08 & -11.79 & -11.16 & -11.53         &-11.59     & \textit{-10.96} \\
        H$_2$O$_2$      & -11.49 & -12.32 & -11.66 & -12.08         &-12.12     & \textit{-11.70}  \\
        H$_2$O          & -12.64 & -13.11 & -12.73 & -12.99         &-13.06     & \textit{-12.62} \\
        CO$_2$          & -13.78 & -14.46 & -13.66 & -14.16         &-14.16     & \textit{-13.77} \\
        CO              & -14.05 & -15.25 & -14.08 & **             &-14.57     & \textit{-14.01} \\
        SO$_2$          & -12.41 & -13.16 & -12.21 & -12.90         &-12.70     & \textit{-12.50}  \\
        ClF             & -12.82 & -13.29 & -12.49 & -13.13         &-13.01     & \textit{-12.77} \\
        CH$_3$Cl        & -11.41 & -11.76 & -11.16 & -11.60$^*$     &-11.66     & \textit{-11.29} \\
        CH$_3$SH        & -9.49  & -9.84  & -9.16  & -9.66          &-9.70      & \textit{-9.44} \\
        SiO             & -11.55 & -12.04 & -11.25 & **             &-11.69     & -11.3  \\
        CS              & -11.45 & -12.55 & -11.34 & **             &-11.83     & -11.33 \\
        HClO            & -11.30 & -11.84 & -11.10 & -11.54$^*$     &-11.59     & -11.12  \\
        \hline
        MAE from exp.               &  0.23  &  0.65  &  0.30  &  0.42  &  0.44 \\
        MAE from $\Delta$CCSD(T)    &        &  0.49  &  0.24  &  0.28  &  0.29 \\
        \hline \hline
    \end{tabular}
\end{table*}

\pagebreak

\bibliography{gw_biblio}
\end{document}